\pdfoutput=1
\documentclass{article}
\usepackage{ijcai20}

\usepackage{times}
\usepackage{soul}
\usepackage{url}
\usepackage[hidelinks]{hyperref}
\usepackage[utf8]{inputenc}
\usepackage[small]{caption}
\usepackage{graphicx}
\usepackage{amsmath}
\usepackage{amsthm}
\usepackage{booktabs}
\usepackage{algorithm}
\usepackage{algorithmic}
\usepackage{multirow}
\usepackage{mathtools}
\usepackage[T1]{fontenc}
\usepackage{microtype}
\usepackage{tikz}

\newcommand{\0}{\phantom0}

\title{Unsatisfiability Proofs for Weight 16 Codewords in Lam's Problem}

\author{
Curtis Bright$^{1,2}$
\and
Kevin K. H. Cheung$^2$\and
Brett Stevens$^2$\and
Ilias Kotsireas$^3$\And
Vijay Ganesh$^1$
\affiliations
$^1$University of Waterloo, Department of Electrical and Computer Engineering\\
$^2$Carleton University, School of Mathematics and Statistics\\
$^3$Wilfrid Laurier University, Department of Physics and Computer Science
}

\begin{document}

\maketitle

\begin{abstract}
In the 1970s and 1980s, searches performed
by L.~Carter, C.~Lam, L.~Thiel, and S.~Swiercz
showed that projective planes of order ten with
weight 16 codewords do not exist.  These searches required highly
specialized and optimized computer programs and required about 2,000 hours
of computing time on mainframe and supermini computers.
In 2011, these searches were verified by D.~Roy using an optimized
C program and 16,000 hours on a cluster of desktop machines.
We performed a verification of these searches by reducing
the problem to the Boolean satisfiability problem (SAT).
Our verification uses the cube-and-conquer SAT solving paradigm,
symmetry breaking techniques using the computer algebra system Maple,
and a result of Carter that there are ten nonisomorphic cases to check.
Our searches completed in about 30 hours on a desktop machine and produced nonexistence proofs
of about 1 terabyte in the DRAT (deletion resolution asymmetric tautology) format.
\end{abstract}

\section{Introduction}

Geometry is one of the oldest branches of mathematics, being first
axiomatically studied by Euclid in the 3rd century BC.
Given a line and a point not on it, Euclid's ``parallel postulate''
implies that there exists exactly one line through the point and
parallel to the given line.
For 2000 years mathematicians tried in vain to prove this axiom but
eventually geometries that did not satisfy the parallel
postulate were discovered.
For example, in the early seventeenth century G. Desargues
studied \emph{projective geometry} where parallel lines do not exist.
Projective geometry became widely studied
in the nineteenth century, leading to the discovery of projective
geometries containing a finite number of points.

Despite a huge amount of study for over 200 years, some basic questions about finite projective
geometries remain open---for example, how many points can a finite projective
plane contain?
It is known~\cite{kaahrstrom2002projective} that this number
must be of the form $n^2+n+1$ for some natural
number~$n$ (known as the \emph{order} of the plane)
and certain orders such as $n=6$ have been ruled out by theoretical arguments.
For every other~$n$ up to ten a finite projective plane of order~$n$ can be
shown to exist through an explicit construction.
No theoretical explanation is known that answers the question
if a projective plane of order ten exists and
answering this question has since become known
as \emph{Lam's problem}.
In the 1970s and 1980s an enormous amount of computing was
used to show that no such plane exists~\cite{lam1991search}.
The computations were based on the existence of codewords in the error-correcting
code generated by a projective plane of order ten.
It was shown~\cite{carter1974existence}
that such a code must contain codewords of weights 15, 16, or~19---but
exhaustive searches
showed that such codewords do not
exist.

Each search required more advanced search techniques
and orders of magnitude more computational power than the previous search---%
the weight~15 search being the easiest and the weight~19 search being the
most challenging.  
In this paper we focus on the weight~16 search that
originally required about 2,000 hours on supercomputers and a VAX-11 supermini machine.
Additionally, in 2011, using an optimized C implementation
the weight~16 search was verified
in 16,000 core hours split across fifteen desktop machines~\cite{roy2011confirmation}.

We provide a reduction of the weight~16 codeword existence problem
to the Boolean satisfiability problem (SAT) and a SAT certification
that the resulting instances are unsatisfiable.
This is done using the
cube-and-conquer SAT solving paradigm~\cite{heule2011cube} and uses functionality from
the computer algebra system Maple for the purposes of symmetry breaking.
See Section~\ref{sec:background} for background on the cube-and-conquer paradigm
and Section~\ref{sec:sat} for a description
of our SAT encoding and symmetry breaking methods.
Our search completed in about 30 hours on a desktop machine,
significantly faster than any previous search.

Furthermore, no previous search was able to
provide any kind of a certificate following a successful completion.
Thus, an independent party had to take on faith that the searches 
did in fact complete.
In contrast, our search produces unsatisfiability certificates
that an independent party can use to verify that our searches were 
successfully run to completion.
The proofs of nonexistence generated by the SAT solver amounted
to about 1 terabyte in the uncompressed DRAT (deletion resolution asymmetric tautology) format~\cite{wetzler2014drat}.
See Section~\ref{sec:results} for details on our implementation and results.

We do not claim our search is a formal verification
because our encoding relies on many mathematical properties that were not derived
in a computer-verifiable form, such as the result
that there are ten nonisomorphic cases that need to be considered~\cite%
{carter1974existence} in addition to the correctness of our encoding
and implementation.
However, we now have a potential method for
producing a formal proof: by formally deriving
our SAT encoding from the projective plane axioms.  This would require expertise
in both projective geometry and a formal proof system and would be a
significant undertaking.  However,
the tools to do this already exist and have been used to formally verify
other results derived using SAT certificates~\cite{cruz2018formally,keller2019}.

\section{Background}\label{sec:background}

We now describe the background necessary to understand the nonexistence
results of this paper, including the method
that we used to solve the SAT instances
and the mathematical background on projective planes
and their symmetry groups that is necessary to understand our SAT reduction.

\paragraph{The cube-and-conquer paradigm.}

The cube-and-conquer paradigm was first developed by Heule, Kullmann,
Wieringa, and Biere~\cite{heule2011cube} for computing
van der Waerden numbers, a notoriously difficult computational
problem from combinatorics.
In recent years the cube-and-conquer method has
been used to resolve long-standing combinatorial problems
such as the Boolean Pythagorean triples problem~\cite{heule2017solving}
and computing the fifth Schur number~\cite{heule2018schur}.

The idea behind the cube-and-conquer method is to split a SAT instance into
subproblems defined by \emph{cubes} (propositional formulae of the form
$l_1\land\dotsb\land l_n$ where $l_i$ are literals).  Each cube defines a
single subproblem---generated by assuming the cube is true---and each subproblem
is then solved or ``conquered'' either in parallel or in sequence.

\paragraph{Projective planes.}

A projective plane is a collection of points and lines that satisfy certain axioms,
for example, in a projective plane any two lines intersect at a unique point.
Finite projective planes can be defined in terms of
\emph{incidence matrices} that have a $1$ in the $(i,j)$th entry exactly when
the $j$th point is on the $i$th line.  In this framework, a projective plane
of order~$n$ is a square $\{0,1\}$-matrix of order $n^2+n+1$ where
any two rows or any two columns intersect exactly once (where two rows or columns
\emph{intersect} when they share a $1$ in the same position).
To avoid degenerate cases we also require that each row contains at least
two zeros or equivalently that each row contains exactly $n+1$ ones.
Two projective planes are said to be \emph{isomorphic} if one can
be transformed into the other via a series of row or column permutations.

Projective planes are known to exist in all orders that are primes or
prime powers and the \emph{prime power conjecture}
is that they exist in no other orders.
Some orders such as $n=6$ have been ruled out on theoretical grounds
making $n=10$ the first uncertain case.
This stimulated a massive computer
search for such a plane~\cite{lam1991search} based on the form such
a plane must have assuming certain codewords exist.
A \emph{codeword} is a $\{0,1\}$-vector in the rowspace (mod 2)
of a $\{0,1\}$-matrix and the \emph{weight} of a codeword is the number
of $1$s that it contains.

\begin{table}
\centering
\begin{tabular}{cccc}
Case & Symmetries & Group Size & Initial Cols. \\
1a & $S_4\wr S_2$ & 1152 & 28 \\
1b & $S_4\times S_4$ & \0576 & 23 \\
1c & $S_4\wr S_2$ & 1152 & 18 \\
2 & $S_4\times S_2$ & \0\048 & 28 \\
3 & $D_8$ & \0\016 & 28 \\
4 & $D_4\times S_2$ & \0\016 & 28 \\
5 & $S_3\times S_2$ & \0\012 & 28 \\
6a & $S_2\times S_2$ & \0\0\04 & 28 \\
6b & $S_2$ & \0\0\02 & 26 \\
6c & $S_2\times S_2$ & \0\0\04 & 24
\end{tabular}
\caption{The ten possible cases for the first eight rows of a projective plane
of order ten generating a weight 16 codeword
and the symmetries in the initial columns (see below).
Here $S_n$ denotes the symmetric group of order~$n!$, $D_n$ denotes
the dihedral group of order~$2n$, and $\wr$ denotes the wreath product.}\label{tbl:cases}
\end{table}

It is known~\cite{carter1974existence,hall1980configurations}
that a projective plane of order ten must generate codewords of weight
15, 16, or~19, thus dramatically shrinking the search space
and naturally splitting the search into three cases.
As shown by~\cite{carter1974existence}, up to isomorphism
there are ten possibilities for the first
eight rows of the planes that generate weight 16 codewords.
Five of these possibilities (cases~2 to~6a in Table~\ref{tbl:cases}) were
eliminated by the searches of~\cite{carter1974existence} and the other five were
eliminated by the searches of~\cite{lam1986nonexistence}.

\paragraph{Incidence matrix structure.}

Carter derived numerous properties that the structure of a projective plane
generating a weight 16 codeword must satisfy.  In particular, the projective plane can be
decomposed into a $3\times2$ grid of submatrices as follows:
\[
\begin{matrix}
& & 16 & 95 \\
\phantom{0}8\!\!\!\! & \multirow{3}{*}{\rlap{$\left(\rule{0pt}{16pt}\right.$}} & 2 & k & \multirow{3}{*}{\llap{$\left.\rule{0pt}{16pt}\right)$}} \\
72\!\!\!\! & & 9 & 8-2k \\
31\!\!\!\! & & 0 & k+3
\end{matrix}
\]

Here the numbers outside the matrix denote the number of rows or columns in that part of the submatrix.
The numbers inside the matrix denote how many $1$s there are in each column in that part of the
submatrix; certain columns depend on a parameter~$k$ that differs between columns.
Additionally, Carter showed that every entry in the first 16 columns
is uniquely specified by the starting case.
We call the columns incident with at least two of the first eight rows the \emph{initial} columns
and the columns incident with at least one of the first eight rows the \emph{inside} columns.
Full starting matrices
for each case are available
at \href{https://uwaterloo.ca/mathcheck/w16}{uwaterloo.ca/mathcheck/w16}.

\paragraph{Symmetry groups.}

A projective plane (or partial projective plane) may be symmetric in nontrivial ways,
in other words, there may exist row or column permutations that fix the entries of the plane.
Such symmetries are important to detect because they can dramatically reduce the search space%
---and therefore the running time---of any search that makes use of them~\cite{aloul2003efficient,heule2015expressing}.

\begin{figure}
\centering
\begin{tikzpicture}[scale=0.33,ultra thin]
\draw [fill=black] (0,0) rectangle (1,-1)
 node[midway,white] {\small1};\draw [fill=black] (1,0) rectangle (2,-1)
 node[midway,white] {\small1};\draw [fill=black] (2,0) rectangle (3,-1)
 node[midway,white] {\small1};\draw [fill=black] (3,0) rectangle (4,-1)
 node[midway,white] {\small1};\draw (4,0) rectangle (5,-1)
 node[midway] {\small0};\draw (5,0) rectangle (6,-1)
 node[midway] {\small0};\draw (6,0) rectangle (7,-1)
 node[midway] {\small0};\draw (7,0) rectangle (8,-1)
 node[midway] {\small0};\draw (8,0) rectangle (9,-1)
 node[midway] {\small0};\draw (9,0) rectangle (10,-1)
 node[midway] {\small0};\draw (10,0) rectangle (11,-1)
 node[midway] {\small0};\draw (11,0) rectangle (12,-1)
 node[midway] {\small0};\draw (12,0) rectangle (13,-1)
 node[midway] {\small0};\draw (13,0) rectangle (14,-1)
 node[midway] {\small0};\draw (14,0) rectangle (15,-1)
 node[midway] {\small0};\draw (15,0) rectangle (16,-1)
 node[midway] {\small0};\draw [fill=black] (16,0) rectangle (17,-1)
 node[midway,white] {\small1};\draw (17,0) rectangle (18,-1)
 node[midway] {\small0};\draw (0,-1) rectangle (1,-2)
 node[midway] {\small0};\draw (1,-1) rectangle (2,-2)
 node[midway] {\small0};\draw (2,-1) rectangle (3,-2)
 node[midway] {\small0};\draw (3,-1) rectangle (4,-2)
 node[midway] {\small0};\draw [fill=black] (4,-1) rectangle (5,-2)
 node[midway,white] {\small1};\draw [fill=black] (5,-1) rectangle (6,-2)
 node[midway,white] {\small1};\draw [fill=black] (6,-1) rectangle (7,-2)
 node[midway,white] {\small1};\draw [fill=black] (7,-1) rectangle (8,-2)
 node[midway,white] {\small1};\draw (8,-1) rectangle (9,-2)
 node[midway] {\small0};\draw (9,-1) rectangle (10,-2)
 node[midway] {\small0};\draw (10,-1) rectangle (11,-2)
 node[midway] {\small0};\draw (11,-1) rectangle (12,-2)
 node[midway] {\small0};\draw (12,-1) rectangle (13,-2)
 node[midway] {\small0};\draw (13,-1) rectangle (14,-2)
 node[midway] {\small0};\draw (14,-1) rectangle (15,-2)
 node[midway] {\small0};\draw (15,-1) rectangle (16,-2)
 node[midway] {\small0};\draw [fill=black] (16,-1) rectangle (17,-2)
 node[midway,white] {\small1};\draw (17,-1) rectangle (18,-2)
 node[midway] {\small0};\draw (0,-2) rectangle (1,-3)
 node[midway] {\small0};\draw (1,-2) rectangle (2,-3)
 node[midway] {\small0};\draw (2,-2) rectangle (3,-3)
 node[midway] {\small0};\draw (3,-2) rectangle (4,-3)
 node[midway] {\small0};\draw (4,-2) rectangle (5,-3)
 node[midway] {\small0};\draw (5,-2) rectangle (6,-3)
 node[midway] {\small0};\draw (6,-2) rectangle (7,-3)
 node[midway] {\small0};\draw (7,-2) rectangle (8,-3)
 node[midway] {\small0};\draw [fill=black] (8,-2) rectangle (9,-3)
 node[midway,white] {\small1};\draw [fill=black] (9,-2) rectangle (10,-3)
 node[midway,white] {\small1};\draw [fill=black] (10,-2) rectangle (11,-3)
 node[midway,white] {\small1};\draw [fill=black] (11,-2) rectangle (12,-3)
 node[midway,white] {\small1};\draw (12,-2) rectangle (13,-3)
 node[midway] {\small0};\draw (13,-2) rectangle (14,-3)
 node[midway] {\small0};\draw (14,-2) rectangle (15,-3)
 node[midway] {\small0};\draw (15,-2) rectangle (16,-3)
 node[midway] {\small0};\draw [fill=black] (16,-2) rectangle (17,-3)
 node[midway,white] {\small1};\draw (17,-2) rectangle (18,-3)
 node[midway] {\small0};\draw (0,-3) rectangle (1,-4)
 node[midway] {\small0};\draw (1,-3) rectangle (2,-4)
 node[midway] {\small0};\draw (2,-3) rectangle (3,-4)
 node[midway] {\small0};\draw (3,-3) rectangle (4,-4)
 node[midway] {\small0};\draw (4,-3) rectangle (5,-4)
 node[midway] {\small0};\draw (5,-3) rectangle (6,-4)
 node[midway] {\small0};\draw (6,-3) rectangle (7,-4)
 node[midway] {\small0};\draw (7,-3) rectangle (8,-4)
 node[midway] {\small0};\draw (8,-3) rectangle (9,-4)
 node[midway] {\small0};\draw (9,-3) rectangle (10,-4)
 node[midway] {\small0};\draw (10,-3) rectangle (11,-4)
 node[midway] {\small0};\draw (11,-3) rectangle (12,-4)
 node[midway] {\small0};\draw [fill=black] (12,-3) rectangle (13,-4)
 node[midway,white] {\small1};\draw [fill=black] (13,-3) rectangle (14,-4)
 node[midway,white] {\small1};\draw [fill=black] (14,-3) rectangle (15,-4)
 node[midway,white] {\small1};\draw [fill=black] (15,-3) rectangle (16,-4)
 node[midway,white] {\small1};\draw [fill=black] (16,-3) rectangle (17,-4)
 node[midway,white] {\small1};\draw (17,-3) rectangle (18,-4)
 node[midway] {\small0};\draw [fill=black] (0,-4) rectangle (1,-5)
 node[midway,white] {\small1};\draw (1,-4) rectangle (2,-5)
 node[midway] {\small0};\draw (2,-4) rectangle (3,-5)
 node[midway] {\small0};\draw (3,-4) rectangle (4,-5)
 node[midway] {\small0};\draw [fill=black] (4,-4) rectangle (5,-5)
 node[midway,white] {\small1};\draw (5,-4) rectangle (6,-5)
 node[midway] {\small0};\draw (6,-4) rectangle (7,-5)
 node[midway] {\small0};\draw (7,-4) rectangle (8,-5)
 node[midway] {\small0};\draw [fill=black] (8,-4) rectangle (9,-5)
 node[midway,white] {\small1};\draw (9,-4) rectangle (10,-5)
 node[midway] {\small0};\draw (10,-4) rectangle (11,-5)
 node[midway] {\small0};\draw (11,-4) rectangle (12,-5)
 node[midway] {\small0};\draw [fill=black] (12,-4) rectangle (13,-5)
 node[midway,white] {\small1};\draw (13,-4) rectangle (14,-5)
 node[midway] {\small0};\draw (14,-4) rectangle (15,-5)
 node[midway] {\small0};\draw (15,-4) rectangle (16,-5)
 node[midway] {\small0};\draw (16,-4) rectangle (17,-5)
 node[midway] {\small0};\draw [fill=black] (17,-4) rectangle (18,-5)
 node[midway,white] {\small1};\draw (0,-5) rectangle (1,-6)
 node[midway] {\small0};\draw [fill=black] (1,-5) rectangle (2,-6)
 node[midway,white] {\small1};\draw (2,-5) rectangle (3,-6)
 node[midway] {\small0};\draw (3,-5) rectangle (4,-6)
 node[midway] {\small0};\draw (4,-5) rectangle (5,-6)
 node[midway] {\small0};\draw [fill=black] (5,-5) rectangle (6,-6)
 node[midway,white] {\small1};\draw (6,-5) rectangle (7,-6)
 node[midway] {\small0};\draw (7,-5) rectangle (8,-6)
 node[midway] {\small0};\draw (8,-5) rectangle (9,-6)
 node[midway] {\small0};\draw [fill=black] (9,-5) rectangle (10,-6)
 node[midway,white] {\small1};\draw (10,-5) rectangle (11,-6)
 node[midway] {\small0};\draw (11,-5) rectangle (12,-6)
 node[midway] {\small0};\draw (12,-5) rectangle (13,-6)
 node[midway] {\small0};\draw [fill=black] (13,-5) rectangle (14,-6)
 node[midway,white] {\small1};\draw (14,-5) rectangle (15,-6)
 node[midway] {\small0};\draw (15,-5) rectangle (16,-6)
 node[midway] {\small0};\draw (16,-5) rectangle (17,-6)
 node[midway] {\small0};\draw [fill=black] (17,-5) rectangle (18,-6)
 node[midway,white] {\small1};\draw (0,-6) rectangle (1,-7)
 node[midway] {\small0};\draw (1,-6) rectangle (2,-7)
 node[midway] {\small0};\draw [fill=black] (2,-6) rectangle (3,-7)
 node[midway,white] {\small1};\draw (3,-6) rectangle (4,-7)
 node[midway] {\small0};\draw (4,-6) rectangle (5,-7)
 node[midway] {\small0};\draw (5,-6) rectangle (6,-7)
 node[midway] {\small0};\draw [fill=black] (6,-6) rectangle (7,-7)
 node[midway,white] {\small1};\draw (7,-6) rectangle (8,-7)
 node[midway] {\small0};\draw (8,-6) rectangle (9,-7)
 node[midway] {\small0};\draw (9,-6) rectangle (10,-7)
 node[midway] {\small0};\draw [fill=black] (10,-6) rectangle (11,-7)
 node[midway,white] {\small1};\draw (11,-6) rectangle (12,-7)
 node[midway] {\small0};\draw (12,-6) rectangle (13,-7)
 node[midway] {\small0};\draw (13,-6) rectangle (14,-7)
 node[midway] {\small0};\draw [fill=black] (14,-6) rectangle (15,-7)
 node[midway,white] {\small1};\draw (15,-6) rectangle (16,-7)
 node[midway] {\small0};\draw (16,-6) rectangle (17,-7)
 node[midway] {\small0};\draw [fill=black] (17,-6) rectangle (18,-7)
 node[midway,white] {\small1};\draw (0,-7) rectangle (1,-8)
 node[midway] {\small0};\draw (1,-7) rectangle (2,-8)
 node[midway] {\small0};\draw (2,-7) rectangle (3,-8)
 node[midway] {\small0};\draw [fill=black] (3,-7) rectangle (4,-8)
 node[midway,white] {\small1};\draw (4,-7) rectangle (5,-8)
 node[midway] {\small0};\draw (5,-7) rectangle (6,-8)
 node[midway] {\small0};\draw (6,-7) rectangle (7,-8)
 node[midway] {\small0};\draw [fill=black] (7,-7) rectangle (8,-8)
 node[midway,white] {\small1};\draw (8,-7) rectangle (9,-8)
 node[midway] {\small0};\draw (9,-7) rectangle (10,-8)
 node[midway] {\small0};\draw (10,-7) rectangle (11,-8)
 node[midway] {\small0};\draw [fill=black] (11,-7) rectangle (12,-8)
 node[midway,white] {\small1};\draw (12,-7) rectangle (13,-8)
 node[midway] {\small0};\draw (13,-7) rectangle (14,-8)
 node[midway] {\small0};\draw (14,-7) rectangle (15,-8)
 node[midway] {\small0};\draw [fill=black] (15,-7) rectangle (16,-8)
 node[midway,white] {\small1};\draw (16,-7) rectangle (17,-8)
 node[midway] {\small0};\draw [fill=black] (17,-7) rectangle (18,-8)
 node[midway,white] {\small1};\end{tikzpicture}
\caption{The upper-left $8\times18$ submatrix from case 1c.
}\label{fig:tetrahedrons}
\end{figure}

For example, Figure~\ref{fig:tetrahedrons} shows the
\emph{initial configuration}
(the first eight rows and initial columns) from case 1c.
This matrix is fixed by the permutation that swaps the first two
rows and column~$k$ with column~$k+4$ for $1\leq k\leq 4$.
The set of all row and column permutations that fix the entries of a matrix
forms a group known as the \emph{symmetry group} of the matrix.

In the matrix of Figure~\ref{fig:tetrahedrons}
any permutation of the first four rows, any permutation of the last four rows,
and the permutation that swaps row~$i$ and row~$i+4$ for $1\leq i\leq4$ occur
(with appropriate column permutations) in the symmetry group.
The size of this permutation group is $4!^2\cdot2=1152$ and the group is isomorphic
to the group of symmetries of a pair of tetrahedrons.
Up to isomorphism, the symmetry groups for each of the ten possible
initial configurations are given in Table~\ref{tbl:cases}.

\section{SAT Encoding}\label{sec:sat}

We now describe the encoding that we use to prove that
projective planes of order ten with weight~16 codewords do not exist.  Although
SAT solvers have been widely used in searching for combinatorial
designs~\cite{zhang2009combinatorial} they have only recently
been applied to projective planes~\cite{bright2020nonexistence}.
Our encoding that an incidence matrix defines a projective plane
is described in Section~\ref{subsec:incidence}, although we do not encode
the entire matrix---we choose the
number of rows and columns to use based on the structure
of the starting cases and in an attempt to minimize the number of
variables and constraints necessary.
Sections~\ref{subsec:colsym} and~\ref{subsec:rowsym} describe the
methods that we use to reduce the symmetries present in the search.

\subsection{Incidence Constraints}\label{subsec:incidence}

First, we describe how we encode the property that the incidence matrix defined
by Boolean variables $p_{i,k}$ (representing that the $(i,k)$th entry of
the matrix contains a~$1$) satisfy the properties of a projective plane.  In particular,
we encode the property that any two rows or columns of the projective plane intersect
exactly once by encoding that any two rows or columns intersect (a) at most once
and (b) at least once.
In fact, we only encode intersections occurring in the first 80 rows---%
our searches find no satisfying assignments of even this strictly smaller
set of constraints.

For (a), the constraints
$\bigwedge_{k,l=1}^{111} (\lnot p_{i,k}\lor\lnot p_{i,l}\lor\lnot p_{j,k}\lor\lnot p_{j,l})$
say that row $i$ and row $j$ do not intersect twice.
We include these constraints
for each distinct pair of indices $(i,j)$ with $1\leq i,j\leq80$
and ignore the constraints where $k$ or~$l$ is
larger than the last column used (which varied between cases).

For (b), the constraint
$\bigvee_{k=1}^{111}(p_{i,k}\land p_{j,k})$
says that row $i$ and row $j$ intersect at least once.
However, this constraint is not in conjunctive normal form so it
can't directly be used with a typical SAT solver.
Instead, we use this constraint in the form
$\bigvee_{k\in S(i)}p_{j,k}$
where $S(i)$ is the set of indices $k$
such that $p_{i,k}$ is true.  The first eight rows of the projective
plane are completely known beforehand in each starting case, so $S(i)$
is well-defined for $1\leq i\leq 8$.
We include these constraints for
all $1\leq i\leq 8$ and $9\leq j\leq 80$.

Similarly, we include constraints that say that column $k$ and column $l$
intersect at least once.
These constraints are of the form $\bigvee_{i\in T(k)}p_{i,l}$
where $T(k)$ is the set of indices~$i$ such that $p_{i,k}$ is true.
We used these constraints for all~$k$ between 1 and 16
and for all~$l>16$ up to the last column used.

In order to use these constraints we require at least 80 rows
and all inside columns because the indices in the set $S(i)$
for $1\leq i\leq 8$ are from the inside columns,
and the indices in the set $T(k)$ for $1\leq k\leq 16$
are from the first 80 rows.
Experimentally we find that \emph{only} using inside columns produces
satisfiable instances---to make all instances unsatisfiable we use
five additional `outside' columns
(selected to be incident to a row needing another five points to
contain eleven $1$s).

\subsection{Breaking Column Symmetries}\label{subsec:colsym}

\begin{figure}
\input 6c.tikz
\caption{The upper-left $8\times62$ submatrix in case 6c
where black squares represent $1$s and white squares represent $0$s.
The first 24 columns form the initial configuration and without
loss of generality the remaining inside columns are taken in lexicographic order.
}\label{fig:tentc}
\end{figure}

Consider the starting matrix shown in Figure~\ref{fig:tentc}
(the first eight rows of case 6c).
This matrix has a symmetry group containing $5!^6\cdot4!^2\cdot2^2$ permutations.
The factor of~$2^2$ arises from symmetries that involve row permutations
and we discuss how we handle those in Section~\ref{subsec:rowsym}.  The larger
$5!^6\cdot4!^2$ factor arises from column permutations of the last 38 columns
which can be split into six blocks of five columns and two blocks of four columns.
We break these symmetries by enforcing a lexicographic ordering on certain
submatrices of the columns.

For example, consider the final block of case 6c
(the last four columns shown in Figure~\ref{fig:tentc}).
By the definition of a projective plane each of these columns must intersect
the first column somewhere in the matrix.  An examination of the full matrix
shows there are exactly six rows where this intersection could happen.
Figure~\ref{fig:lex} shows the case when the rows of the matrix are
ordered so that the six rows occur adjacent
to each other---so that the intersections must occur in a $6\times4$ submatrix.
We break the column symmetries
by lexicographically ordering the columns in this submatrix.

Each row of the submatrix contains at most a single~$1$
or we would have a pair of columns that intersect each other twice.
Thus, lexicographically ordering the columns of the submatrix
ensures that a~$1$ cannot be in the upper-right or lower-left corners
of the submatrix as displayed by the $0$s in Figure~\ref{fig:lex}.

\begin{figure}
\centering\begin{tikzpicture}[scale=0.33,ultra thin]
\draw [fill=black,draw=black,shift={(0,1.75)}] (2,-1) rectangle (3,-2) node[midway,white] {\small1};
\draw [fill=black,draw=black,shift={(0,1.75)}] (3,-1) rectangle (4,-2) node[midway,white] {\small1};
\draw [fill=black,draw=black,shift={(0,1.75)}] (4,-1) rectangle (5,-2) node[midway,white] {\small1};
\draw [fill=black,draw=black,shift={(0,1.75)}] (5,-1) rectangle (6,-2) node[midway,white] {\small1};
\draw [fill=black,draw=black,shift={(-0.75,0)}] (0,-2) rectangle (1,-3) node[midway,white] {\small1};
\draw (2,-2) rectangle (3,-3);
\draw (3,-2) rectangle (4,-3) node[midway] {\small0};
\draw (4,-2) rectangle (5,-3) node[midway] {\small0};
\draw (5,-2) rectangle (6,-3) node[midway] {\small0};
\draw [fill=black,draw=black,shift={(-0.75,0)}] (0,-3) rectangle (1,-4) node[midway,white] {\small1};
\draw (2,-3) rectangle (3,-4);
\draw (3,-3) rectangle (4,-4);
\draw (4,-3) rectangle (5,-4) node[midway] {\small0};
\draw (5,-3) rectangle (6,-4) node[midway] {\small0};
\draw [fill=black,draw=black,shift={(-0.75,0)}] (0,-4) rectangle (1,-5) node[midway,white] {\small1};
\draw (2,-4) rectangle (3,-5);
\draw (3,-4) rectangle (4,-5);
\draw (4,-4) rectangle (5,-5) node[midway] {\small$y$};
\draw (5,-4) rectangle (6,-5) node[midway] {\small0};
\draw [fill=black,draw=black,shift={(-0.75,0)}] (0,-5) rectangle (1,-6) node[midway,white] {\small1};
\draw (2,-5) rectangle (3,-6) node[midway] {\small0};
\draw (3,-5) rectangle (4,-6) node[midway] {\small$x$};
\draw (4,-5) rectangle (5,-6);
\draw (5,-5) rectangle (6,-6);
\draw [fill=black,draw=black,shift={(-0.75,0)}] (0,-6) rectangle (1,-7) node[midway,white] {\small1};
\draw (2,-6) rectangle (3,-7) node[midway] {\small0};
\draw (3,-6) rectangle (4,-7) node[midway] {\small0};
\draw (4,-6) rectangle (5,-7);
\draw (5,-6) rectangle (6,-7);
\draw [fill=black,draw=black,shift={(-0.75,0)}] (0,-7) rectangle (1,-8) node[midway,white] {\small1};
\draw (2,-7) rectangle (3,-8) node[midway] {\small0};
\draw (3,-7) rectangle (4,-8) node[midway] {\small0};
\draw (4,-7) rectangle (5,-8) node[midway] {\small0};
\draw (5,-7) rectangle (6,-8);
\node at (4,-0.8) {\small$\vdots$};
\node at (1.25,-5) {\small$\cdots$};
\end{tikzpicture}
\caption{A $6\times4$ submatrix demonstrating our column symmetry breaking encoding.
The~$0$s that appear are known under
the assumption that the submatrix's columns are lexicographically ordered.
This also implies that the entry marked~$y$ is~$0$
if the entry marked~$x$ contains a~$1$.
}\label{fig:lex}
\end{figure}

Similarly, each column of the submatrix contains at most a single~$1$
(in fact it must contain exactly a single~$1$ since
the first column must intersect each of the other columns in the submatrix).
Consider the entry marked~$x$ in Figure~\ref{fig:lex}.  If this entry
contains a~$1$ then all entries in the next column that are above it
must be~$0$ for the columns to be correctly ordered.

Thus, considering the variables labelled~$x$ and~$y$
in Figure~\ref{fig:lex}, we include the clause
$\lnot x\lor\lnot y$ in our encoding.
We include similar clauses for each unknown entry in the submatrix.
The remaining satisfying assignments are exactly those whose columns
are in lexicographic order, therefore enforcing a unique ordering on columns
which are otherwise identical.  
By using this method on each block of column symmetries
this decreases the search space size
by a factor of $5!^6\cdot4!^2\approx10^{15}$
in the case of Figure~\ref{fig:tentc}.
Each block is independent of
the others and will have its symmetries broken in a different submatrix.

\subsection{Breaking Initial Symmetries}\label{subsec:rowsym}

In Section~\ref{subsec:colsym} we described how we break most of the symmetries
in our instances.  However, it remains to break the initial symmetries involving
\emph{both} row and column permutations (i.e., those described in Table~\ref{tbl:cases}).
We used two methods of breaking these symmetries.  One method was particularly
effective when the size of the symmetry group was not too large and was used
in the cases~2 to~6c.  The second method took advantage of the specific form
of the symmetry group of the cases~1a--c.

\paragraph{Lex method.}

In addition to the column symmetries of the last 38 columns of the matrix
in Figure~\ref{fig:tentc} there are an additional
two generators $\varphi_1$ and $\varphi_2$ of the symmetry group of this matrix.
These generators involve both
row and column permutations.  In cycle notation the row permutations
are $(1,5)(2,6)(3,4)$ and $(1,2)(3,4)(5,6)(7,8)$ and
the column permutations are completely determined by these row permutations.
For example, after swapping rows~1 and~5 and rows~2 and~6, the
first column (containing~$1$s in rows~1 and~2) becomes
the 14th column (containing~$1$s in rows~5 and~6), so~$\varphi_1$ must send
column~1 to column~14.

Similarly, permutations of the first eight rows can be extended
to permutations of the entries in the submatrix~$P$
consisting of the first 80 rows and the initial columns---the extension
uniquely determined by the restriction that the initially
known entries remain fixed.
We now consider the action of these permutations on the submatrix~$P$.
For example, in case~6c
the first 24 columns form the initial columns and
the lex symmetry breaking method focuses on the unknown
entries in the upper-left $80\times24$ submatrix~$P$.

In particular, we fix an ordering on the variables in the
submatrix $P$ whose values are undetermined.  For example, one possible ordering
of these variables is left-to-right and top-to-bottom, i.e.,
\[ L_P \coloneqq [p_{9,17},\, p_{9,18},\, p_{9,19},\, \dotsc,\, p_{80,22},\, p_{80,23},\, p_{80,24}] . \]
Under the symmetry $\varphi_1$ this ordering becomes
\[ [p_{56,19},\, p_{56,22},\, p_{56,17},\, \dotsc,\, p_{17,18},\, p_{17,20},\, p_{17,24}] \]
which we denote by $\varphi_1(L_P)$.
If $P$ is a $80\times24$ partial projective plane then $\varphi_1(P)$ is also a $80\times24$ partial projective plane.
It follows that any partial projective plane of this form
can be transformed into an isomorphic partial projective plane with
$L_P \leq_\text{lex} \varphi_1(L_P)$.
Similarly, up to equivalence we can assume that
\[ L_P \leq_{\text{lex}} \varphi(L_P) \label{eq:star}\tag{$*$} \]
where $\varphi$ is any symmetry of $P$.
In the example of Figure~\ref{fig:tentc}
we take $\varphi$ to be $\varphi_1$,
$\varphi_2$, and $\varphi_1\circ\varphi_2$ (there is no need to
take~$\varphi$ to be the identity symmetry since then~\eqref{eq:star} is trivial).

It remains to describe how we encode the lexicographic constraint~\eqref{eq:star}
in conjunctive normal form.
We express the general lexicographic constraint
$[x_1,\dotsc,x_n]\leq_\text{lex}[y_1,\dotsc,y_n]$
using $3n-2$ clauses and $n-1$
new variables~\cite{knuth2015art}.
Denoting the new variables by $a_1$, $\dotsc$, $a_{n-1}$,
the clauses are
\[ \lnot x_k\lor y_k\lor\lnot a_{k-1},\,\,\,\lnot x_k\lor a_k\lor\lnot a_{k-1},\,\,\,y_k\lor a_k\lor\lnot a_{k-1} \]
for $k=1$, $\dotsc$, $n-1$ (with $\lnot a_0$ omitted) along with the clause $\lnot x_n\lor y_n\lor\lnot a_{n-1}$.

In practice this method breaks almost all of the symmetries
of the search space that exist in the initial columns, though it is not
guaranteed to break them all.
This is because for certain instantiations of the variables of
$P$ it may be the case that~\eqref{eq:star}
does not remove any isomorphic solutions.
This occurs when $P$ is unchanged under all symmetries $\varphi$ of the symmetry group
(i.e., when $L_P=\varphi(L_P)$ for all symmetries~$\varphi$).
However, this did not occur very often in practice.

Similar to the work~\cite{crawford1996symmetry} this method uses constraints of the form~\eqref{eq:star}
for every nontrivial symmetry.  Because of this, it worked best
when the symmetry group was relatively small.  We used this symmetry breaking
method on the starting cases 2--6c.

\paragraph{Block method.}

Consider the starting matrix given in Figure~\ref{fig:tetrahedrons} (case 1c).
Since each row contains eleven~$1$s in total and there are five~$1$s per row in the initial
columns, there are another six~$1$s in each row.
Furthermore, since the rows are already pairwise intersecting in the initial columns, none
of the rows intersect following the initial columns.
In other words, each row contains six $1$s on columns
that are not incident with the other rows.
We call the columns that contain those six $1$s a \emph{block} of $1$s.

Without loss of generality we may assume that the first block occurs on
columns 19--24, the second block occurs on columns 25--30, the third
block on columns 31--36, etc.  By exhaustive search, we find that there
are 49,472 solutions of the SAT instance that uses the first 80 rows, the columns
up to and including the first block,
and the lexicographic column ordering clauses described
in Section~\ref{subsec:colsym}.

Our symmetry breaking method now uses the following properties of the symmetry
group of case 1c:
\begin{description}
\item[\rm1.] The symmetry group fixing the first row (i.e., the symmetries
that do not move the first block) is isomorphic to $S_4\times S_3$.
It has a set of generators $\{\varphi_2,\varphi_3,\varphi_5,\varphi_6,\varphi_7\}$ where $\varphi_i$
is a permutation that swaps rows $i$ and $i+1$ (and appropriate columns)
but fixes each of the other first eight rows.
\item[\rm2.] The action of the symmetry group on the set of blocks is transitive,
i.e., for each pair of blocks there exists a symmetry that sends one block
to another block.  For example, $\varphi_1$ sends block~1 to block~2.
\end{description}
We use the first property to show that of the 49,472 possibilities for the
first block, only 469 possibilities are nonisomorphic under the symmetries
that fix the first row.  These numbers agree with those reported by~\cite{lam1986nonexistence}.
This splits the search into 469 distinct cases, one for each nonequivalent
possibility of the first block.  Additionally, we now use the second property to
remove further symmetries within these cases.

We label each of the 49,472 possibilities for the first block with a label
between~1 and~469 and let $t$ be the labelling function that when given an instantiation
of the first block returns its label.
Furthermore, let $B_i$ denote the submatrix consisting of the $i$th block
where $1\leq i\leq 8$ and let $\psi_i$ denote a symmetry that sends block~$i$
to block~$1$ where $2\leq i\leq8$.
With $\varphi_i$ defined as above, we can take $\psi_2\coloneqq\varphi_1$,
$\psi_5\coloneqq(1,5)(2,6)(3,7)(4,8)$ (as row permutations in cycle notation),
and $\psi_i\coloneqq\varphi_{i-1}\circ\psi_{i-1}$ for all other $i$.
We extend the labelling function~$t$ by giving an instantiation~$B$ of an
arbitrary block~$B_i$ the same label as $\psi_i(B)$.

Up to isomorphism we can assume that
any partial projective plane in case 1c must satisfy
\[ t(B_1) \leq t(B_i) \quad\text{for}\quad 2\leq i\leq 8 . \label{eq:star2}\tag{$**$} \]
In other words, we can assume that the first block has the minimum label of all the
blocks 1--8.  For suppose a partial projective plane did not
satisfy this condition: then there must be an index $m$ such that
block~$m$ has the minimum label.  Applying $\psi_m$ to this
partial projective plane permutes the block labels and
produces an isomorphic partial projective plane such that
block~1 has the minimum label.

To encode the constraint~\eqref{eq:star2} in our SAT instances we use a series of blocking
clauses.  In each SAT instance the left-hand side of~\eqref{eq:star2} is known in advance,
since each
instance contains a fixed instantiation of the first block.
In the first SAT instance the label of the first block is~1.  In this case~\eqref{eq:star2}
is trivial and does not block any solutions.

In the second SAT instance we need to block all solutions where $t(B_i)=1$ for $2\leq i\leq 8$.
To do this, suppose $B$ is an instantiation of the first block that is labelled~1.  We generate
$\psi^{-1}_i(\varphi(B))$ for all $2\leq i\leq 8$ and all $\varphi$ in the symmetry group
fixing the first line.  This gives us an explicit collection of instantiated blocks
that we want to ignore.  If $B'$ is one of these blocks
and $B'\models p$ means that variable $p$ is assigned true in the assignment defined by $B'$ then
the clause $\bigvee_{B'\models p}\lnot p$
prevents $B'$ from occurring in the solution of the SAT instance.  We include
such clauses in the SAT instance for all $B'$ of the form $\psi^{-1}_i(\varphi(B))$.
Similarly, in the $k$th SAT instance we include clauses of this form for all $B'$ of the
form $\psi^{-1}_i(\varphi(B))$ where~$B$ is an instantiation of the first block whose
label is strictly less than~$k$.

The above description specifically applies to case~1c, but cases~1a and~1b can be
handled in a similar way.  In case~1a the main difference is that each block consists
of seven columns and some columns are shared between blocks.  However,
the same method applies because the two properties of the symmetry group that we
used in 1c also hold in this case (cases~1a and~1c have an isomorphic symmetry group).

In case 1a we find 21,408 solutions of the SAT instance using the first
80 rows and the columns up to and including the first block (the block containing
$1$s on the first row).
Using the symmetries that fix the
first row we find that only 275 of the 21,408 solutions are nonisomorphic,
thus naturally splitting the problem into 275 SAT instances
that we solve in the same way as we solve the instances from case~1c.

We solve case 1b in a similar way, but in this case four of the blocks
contain seven columns and the other four blocks contain six columns.
We order the blocks such that the first four blocks contain seven columns
and the last four blocks contain six columns.  The first block coincides
with the first block in case~1a and the symmetry group that fixes the
first row is the same as in case~1a, so we also have 275 nonisomorphic
solutions of the first block in case~1b.

In case~1b the symmetry group does not act transitively on the blocks because
there are no symmetries that send blocks with seven columns to blocks with
six columns.  However, the symmetry group does act transitively
on the first four blocks.  Thus we use the same symmetry breaking condition
given in~\eqref{eq:star2} except replacing the condition on~$i$ with
$2\leq i\leq 4$.

\section{Implementation and Results}\label{sec:results}

The ten starting matrices were constructed following~\cite{carter1974existence}
who provides the first 8 rows and 16 columns of each case.
The symmetry groups in each case were computed using the computer algebra system
Maple~2019 which uses the library nauty~\cite{mckay2014practical}.
A complete collection of the row and column
permutations in each group were saved so that they could be later used in the
symmetry breaking method.  A Python script of about 250 lines was written to
generate the constraints~(a) and~(b) from Section~\ref{subsec:incidence},
the column symmetry breaking constraints from Section~\ref{subsec:colsym},
and for cases 2--6c the symmetry breaking constraints from the lex method in Section~\ref{subsec:rowsym}.
As input it took the case for which to generate constraints and the number
of rows and columns to use (though we used 80 rows in all cases).  The script required
the starting matrix as well as the permutations necessary for breaking
the initial symmetries to be explicitly provided.

\paragraph{Cases 2--6c.}

For these cases the lex symmetry breaking method described in
Section~\ref{subsec:rowsym} was used.  Our Python script was used to generate
a single SAT instance for each case using all inside columns and
an additional five columns.
The cube-and-conquer SAT solving
paradigm~\cite{heule2017solving} was then used to show that each instance was unsatisfiable.
The preprocessor of the SAT solver
Lingeling~\cite{Biere-SAT-Competition-2017-solvers} (at optimization level 5)
simplified the instances prior
to cubing, the lookahead SAT solver March\_cu~\cite{heule2011cube} generated the cubes,
and an incremental version of MapleSAT~\cite{liang2018empirical}
solved the instances and generated proofs of unsatisfiability.

In each case the proofs from the simplification and conquering steps
were combined into a single proof by concatenation
and the combined proof was verified using the proof checker DRAT-trim~\cite{wetzler2014drat}
as well as the GRAT toolchain~\cite{lammich2020efficient}.
These cases took about 10 total hours to generate seven proofs whose size together
totalled about 300 gigabytes in the plain uncompressed DRAT format.

\paragraph{Cases 1a--c.}

In these cases the block symmetry breaking method described in Section~\ref{subsec:rowsym}
was used.
To begin, we generated SAT instances without the initial symmetry breaking clauses
and using the columns up to and including the first block.
In order to find all \emph{nonisomorphic} solutions of the first block
we use a ``programmatic'' (see \cite{ganesh2012lynx}) version of the MapleSAT
solver~\cite{bright2018enumeration}.
Whenever a solution is found we record the solution and then
programmatically learn clauses that
block the solution in addition to all solutions isomorphic to the
found solution (using the symmetry group fixing the first block as computed
by Maple).  In this way, only the nonisomorphic solutions are recorded.

For every nonisomorphic solution a SAT instance is generated (by another Python script)
that includes unit clauses completely specifying the first block and the
symmetry breaking clauses from the block method in Section~\ref{subsec:rowsym}.
The cube-and-conquer method then solves these instances similar
to how the previous cases are solved, though cases 1a--c generated
1019 individual proofs (469 in case~1c and 275 for each
case~1a and~1b).  The fact that these cases are exhaustive relies
on nontrivial symmetry group computations, so we did not attempt to combine these
proofs together.

These instances use five outside columns and all inside columns
with the exception of the noninitial columns in block~5 (in cases 1a--b)
and blocks~2 and~3 (in case 1c)---removing these columns
generates a strictly smaller set of constraints that increases the
performance of the SAT solver.  Similarly,
\cite{lam1986nonexistence} ignored the columns in blocks~2 and~3 of case 1c
and found no solutions.  We verified their result, though
this depends on the choice of representatives chosen for the nonisomorphic
solutions of the first block---some choices of the representatives do in fact lead
to satisfiable instances.

Interestingly, the performance in cases 1a and 1c improves
if the initial symmetry breaking clauses are left out of the instances
provided to the cubing solver,
as the cubes generated in this manner are still effective in the conquering phase.
In case~1b this causes some cubes to have dramatically unbalanced
difficulties, so in this case we include
the initial symmetry breaking clauses in the cubing instances.

Following~\cite{lam1986nonexistence},
the nonisomorphic solutions of the first block are ordered in increasing size of
their stabilizer group.  In other words, the solutions
that are isomorphic to many other solutions are given small labels and the solutions
that are isomorphic to few other solutions are given large labels.

\paragraph{Results.}

\begin{table}
\centering
\begin{tabular}{ccccccccc}
Case & \0Cubes & Cubing Time & Time (h) & Proof size \\
1a & \0222310 & \04.9\% & \02.61 & \082.0G \\
1b & \0379969 & 15.1\% & \02.58 & \089.1G \\
1c & 1300912 & \03.6\% & 13.98 & 551.5G \\
2  & \0\0\04486 & \02.0\% & \01.82 & \018.4G \\
3  & \0\036388 & \07.1\% & \01.39 & \037.6G \\
4  & \0\0\01529 & \08.8\% & \00.03 & \0\01.0G \\
5  & \0\0\04582 & \01.4\% & \00.42 & \0\09.7G \\
6a & \0\0\08109 & \01.9\% & \00.53 & \013.3G \\
6b & \0\091448 & \03.6\% & \02.21 & \076.3G \\
6c & \0224942 & \04.6\% & \03.26 & 141.5G 
\end{tabular}
\caption{Summary of the results of our implementation
applied to all weight 16 cases.
}\label{tbl:results}
\end{table}

The computations were run on a desktop machine with an
Intel i7 CPU at 2.7 GHz.
A summary of our results is presented in Table~\ref{tbl:results}.
In particular, this table specifies the total number of cubes used
in each case, the proportion of the running time spent
running the cubing solver, the total running time (in hours) of the SAT solvers,
and the size of the proofs produced.
The scripts used to generate and solve the SAT instances are available
at \href{https://uwaterloo.ca/mathcheck}{uwaterloo.ca/mathcheck}
and the proofs are archived at
\href{https://zenodo.org/record/3767062}{zenodo.org/record/3767062}.

The amount of cubing was controlled by the cubing cutoff \mbox{\kern1sp\texttt{-n}}
parameter of March\_cu (replacing the default heuristic cutoff).
This cutoff method stops cubing once the number of free variables drops below
the given bound.  We attempted to perform an amount of cubing that would
minimize the total solve time of each case, but we do not claim we used
an exactly optimal amount.
Typically starting with a cubing cutoff bound equal to about
75\% of the free variables in the instance worked well and we tuned the
bound higher or lower based on the solver performance.
The cubing process was done once for each SAT instance
and the cubes were incrementally solved by MapleSAT
(which was used because it performed better than the default cube-and-conquer
solvers).

\section{Conclusion and Future Work}\label{sec:conclusion}

We have provided a SAT certification that there exist
no projective planes of order ten generating weight~16 codewords.
This verifies the original searches of \cite{carter1974existence,lam1986nonexistence},
as well as the verification of \cite{roy2011confirmation} that was
based on an equivalent but different way of decomposing the search.
The previous searches relied on highly optimized computer programs
and special-purpose search algorithms.  In contrast, our search used
widely available and well-tested SAT solvers, computer algebra systems,
and proof verifiers.  Our search produced
the first nonexistence certificates for this problem that can be
checked by a third party.

Furthermore, our search is the fastest known verification of this result.
The search of~\cite{carter1974existence} that used about 140 hours on
supercomputers was verified in about 4 hours, and the search of~\cite{lam1986nonexistence}
that used about 2000 hours on a VAX-11 was verified in about 25 hours.
This is in large part due to the increase in computation power
available today---however, the verification of~\cite{roy2011confirmation}
which used modern AMD CPUs running at 2.4 GHz required 16,000 hours.

A similar SAT encoding has been used to show the nonexistence of weight~15
codewords in a few minutes---and under 10 seconds
using programmatic symmetry breaking~\cite{bright2020nonexistence}.
The next challenge is to show the nonexistence of weight~19 codewords.
We believe our approach will be useful in this
search but it will likely require alternative symmetry breaking
methods specifically tailored to the weight~19 search.

\section*{Acknowledgments}
We thank the reviewers for their detailed
comments that improved the clarity of this article
and we thank the open-access repository Zenodo for archiving our proofs.

\bibliographystyle{named}
\bibliography{unsatproofs-w16}

\end{document}